\documentstyle[11pt]{article} 
\begin{document}
\begin{titlepage}
\title{\bf An Explicit Construction of  Self-dual 2-forms
in Eight Dimensions}
\date{ }
\author{ {\bf Ay\c{s}e H\"{u}meyra Bilge}\\{\small
Department of Mathematics, Institute for Basic Sciences}\\
{\small TUBITAK Marmara Research Center}\\
{\small Gebze-Kocaeli, Turkey}\\
{\footnotesize e.mail : bilge@yunus.mam.tubitak.gov.tr}\\
{\bf Tekin Dereli}\\{\small Department of Mathematics}\\
{\small Middle East Technical University}\\{\small Ankara, Turkey}
\\{\footnotesize e.mail : tdereli@rorqual.cc.metu.edu.tr}\\
{\bf \c{S}ahin Ko\c{c}ak}\\{\small Department of Mathematics}\\
{\small Anadolu University}\\{\small Eski\c{s}ehir, Turkey}\\
{\footnotesize e.mail : d170@tranavm1}}
\maketitle
\begin{abstract}
The geometry of self-dual 2-forms in 2n dimensional spaces is
studied. These  2-forms determine a $n^2-n+1$
dimensional manifold ${\cal S}_{2n}$. We prove that ${\cal S}_{2n}$
has only one-dimensional linear
submanifolds for n odd. In eight dimensions
the self-dual forms of Corrigan et al constitute a seven dimensional
linear subspace of  ${\cal S}_{8}$ among many other equally interesting
linear subspaces.
\end{abstract}
\end{titlepage}
\noindent {\bf 1. Introduction}
\vskip 3mm
The concept of self-duality of a 2-form in four
dimensions is generalised to any higher even dimensional
space in our previous paper ${}^{[1]}$.
We recall here that self-duality can be defined as an eigenvalue
criterion in the following way. ( Here we adopt a different
terminology, and use self-dual rather than strongly self-dual as
it is used in Ref.[1] )
Suppose $F$ is a real 2-form in 2n dimensions,
and let $\Omega$ be the corresponding $2n \times 2n$
skew-symmetric matrix with respect to some local orthonormal basis.
Then by a change of basis, $\Omega$ can be brought to the
block-diagonal form
$$\left ( \begin{array}{ccccccc}
0&\lambda_{1}& & & & & \\
-\lambda_{1}&0& & & & & \\ & &. & & & & \\ & & &. & & & \\ & & & &.& &  \\
& & & &  & 0&\lambda_{n}\\ & & & &  &- \lambda_{n}&0
\end{array} \right )
$$
where $\lambda_{1},...,\lambda_{n}$ are the eigenvalues of $\Omega$.
The 2-form $F$ is called   self-dual or anti-self-dual provided
the absolute values of the eigenvalues are all equal , that is
$$|\lambda_{1}|=|\lambda_{2}|= \dots =|\lambda_{n}|.$$
To distinguish between the two cases, orientation must be taken into account.
We define $F$ to be self-dual, if $\Omega$ can be brought with respect
to an orientation-preserving basis change to the above block-diagonal
form such that $ \lambda_{1}=\lambda_{2}= \dots = \lambda_{n}$.
Similarly, we define $F$ to be anti-self-dual, if $\Omega$
can be brought to the same form  by an orientation-reversing basis
change.
It is not difficult to check that in dimension D=4, the above definition
coincides with the usual definition of self-duality in the Hodge sense.
We studied in our previous work self-dual 2-forms in D=8 and have shown that

i) Yang-Mills 2-forms satisfying a certain set of 21 linear equations,
first derived by Corrigan, Devchand, Fairlie, Nuyts (CDFN) ${}^{[2]}$
by other means, are  self-dual in the above sense.

ii) Each self-dual 2-form $F$, satisfying
$ \ast(F \wedge F) = F \wedge F$
${}^{[3]}$ is self-dual in the above sense.

In this letter, starting from the self-duality condition on eigenvalues
we obtain the CDFN self-dual 2-form.
We also explain the construction of new  families of
self-dual 2-forms.
\newpage
\vskip 0.5cm
\noindent
{\bf 2. The Geometry of Self-dual 2-forms.}
\vskip .2cm

In this section we describe the geometrical structure of  self-dual
2-forms in arbitrary even dimensions. In the following $I$ denotes
an identity
matrix of appropriate dimension.
 \vskip .1cm
 \noindent
{\bf Definition 1.} Let ${\bf A}_{2n}$ be the set of antisymmetric matrices in
$2n$ dimensions. Then ${\cal S}_{2n}=\{ A\in {\bf A}_{2n}
\mid A^2+\lambda^2I=0,\lambda\in {\bf R}, \lambda \ne 0\}$.

Note that if $A\in {\cal S}_{2n}$, and $A^2=0$, then $A=0$, and if
$A\in {\cal S}_{2n}$,
then $\lambda A\in{\cal S}_{2n}$ for $\lambda \ne 0$.

\vskip .2cm

\noindent
\proclaim Proposition 2. The set ${\cal S}_{2n}$ is diffeomorphic to
$\big(O(2n)\cap{\bf A}_{2n}\big)\times {\bf R}^+$.

\noindent
{\it Proof.} Let $A\in{\cal S}_{2n}$ with $A^2+\lambda^2I=0$. Note that
$\lambda^2=-\frac{1}{2n}{\it tr}A^2$.  Define $\kappa=\big[-\frac{1}{2n}{\it
tr}A^2\big]^{1/2}$, $\tilde{A}=\frac{1}{\kappa}A$. Then, $\tilde{A}^2+I=0$,
hence $\tilde{A}\tilde{A}^\dagger=I$. Consider the map
$\varphi:{\cal S}_{2n}\to
\big(O(2n)\cap {\bf A}_{2n}\big)\times {\bf R}^+$ defined by
$\varphi(A)=(\tilde
{A},\kappa)$. The map $\varphi $ is 1-1, onto and differentiable. Its inverse
is given by $(B,\alpha)\to\alpha B$ is also differentiable, hence $\varphi$ is
a diffeomorphism.\quad e.o.p.

\vskip .2cm
\noindent
{\bf Remark 3.}
$ O(2n) \cap {\bf A}_{2n}$ is a fibre bundle over the sphere
$S^{2n-2}$ with fibre $O(2n-2) \cap {\bf A}_{2n-2}$.
(See Steenrod, Ref.{[4]})
\vskip .1cm
For our purposes the following description of ${\cal S}_{2n}$
is more useful.

\proclaim Proposition 4. ${\cal S}_{2n}$ is diffeomorphic to the homogeneous
manifold $\big(O(2n)\times {\bf R}^+\big)/U(n) \times \{ 1 \}$,
and {\it dim}${\cal S}_{2n} =n^2-n+1$.

\vskip .2cm
\noindent
{\it Proof.} Let $G$ be the product group $O(2n)\times {\bf R}^+$,
where ${\bf R}^+$ is
considered as a multiplicative group. $G$ acts on ${\cal S}_{2n}$
by $(P,\alpha)\dot
A=\alpha(P^t A P)$, where $P\in O(2n)$, $\alpha\in {\bf R}^+$,
$A\in{\cal S}_{2n}$, and $t$
indicates the transpose. Since all matrices in ${\cal S}_{2n}$
are conjugate to each
other up to a multiplicative constant, this action is transitive, and actually
any $A\in {\cal S}_{2n}$ can be written as $A=\lambda P^t JP$,
where $\lambda=
\big[-\frac{1}{2n}{\it tr} A^2\big]^{1/2}$, with $P\in O(2n) $ and
$J=\pmatrix{0&I\cr -I&0\cr}$. It can be seen that the isotropy subgroup of $G$
at $J$ is $U(n)$ ${}^{[5]}$ and $G/U(n)$ is diffeomorphic to ${\cal S}_{2n}$
( Ref.{[6]} p.132, Thm.3.62 )  Then {\it dim}${\cal S}_{2n}=dim\big(O(2n)\times
{\bf R}^+/U(n)\big)$ can be easily computed as
{\it dim}${\cal S}=dim O(2n)+1-dimU(n)=(2n^2-n+1)-n^2=n^2-n+1$.
\quad e.o.p.

In particular, in eight dimensions, ${\cal S}_8$ is a 13 dimensional
manifold.

As $O(2n)$ has two connected components ( $SO(2n)$ and $O(2n) \setminus
SO(2n)$), $U(n)$ is connected and $U(n) \subset SO(2n)$,
${\cal S}_{2n}$ has two connected components. One of them
(that contains J) consists of the selfdual forms and the other of the
anti-self-dual forms.
\newpage
\noindent
{\bf 3. Linear submanifolds of ${\cal S}_{8}$ }
\vskip .2cm

The defining equations of the set ${\cal S}_8$ are homogeneous quadratic
polynomial
equations for the components of the curvature 2-form. Thus they correspond to
differential equations quadratic in the first derivative for the connection.
Thus the study of their solutions, hence the study of the moduli
space of  self-dual
connections is rather difficult. Therefore one might hope to
restrict the notion of
self-duality, to the linear submanifolds of ${\cal S}_8 \cup \{0\}$.
But there are plenty of them (at least in ${\cal S}_8$)
and there is no plausible reason to single out a specific one of them.
In Ref.{[1]} we
have shown that the 2-forms satisfying a set of 21 equations proposed by
Corrigan et al belong to  ${\cal S}_8$ and we shall give below a natural
way of arriving at them, but it will depend on a reference form.
Changing the reference form one obtains translates of this 7-dimensional
plane, which in some cases look more pregnant than the original one.

Note that we excluded the zero matrix from ${\cal S}_{2n}$
in our definition in order
to obtain its manifold structure. We denote $\overline{\cal S}_{2n}
={\cal S}_{2n} \cup
\{0\}$. By linearity of the action of $O(2n) $ on ${\cal S}_{2n}$
we obtain the following

\proclaim Lemma 4. Let ${\cal L}$ be a linear submanifold of $\overline{{\cal
S}}_{2n}$. Then
${\cal L}_P=P^t {\cal L} P$, $P\in O(2n)$ is also a linear submanifold of
$\overline{{\cal S}}_{2n}$.

Let $J_o=diag(\epsilon,\epsilon,\epsilon,\epsilon)$, where
$\epsilon=\pmatrix{0&1\cr-1&0\cr}$. Note that any $A\in {\cal S}_8$
is conjugate
to $J_o$, hence any linear subset of $\overline{{\cal S}}_8$ can be realized as
the translate of a linear submanifold containing $J_o$ under conjugation.
Thus without loss of generality we can concentrate on linear subsets
containing $J_o$.

\proclaim Proposition 5. If $A\in {\cal S}_8$ and $(A+J_o)\in {\cal S}_8$,
where
$J_o=diag(\epsilon,\epsilon,\epsilon,\epsilon)$, with
$\epsilon=\pmatrix{0&1\cr-1&0\cr}$, then
$$A=\pmatrix{k\epsilon &r_1S(\alpha) &r_2S(\beta)   &r_3S(\gamma)\cr
        -r_1S(\alpha)   &k\epsilon    &r_3S(\gamma') &-r_2S(\beta')\cr
        -r_2S(\beta)   &-r_3S(\gamma')&k\epsilon    &r_1S(\alpha')\cr
        -r_3S(\gamma)  &r_2S(\beta')  &-r_1S(\alpha')&k\epsilon\cr}$$
where $k\in R$, $r_1$, $r_2$, $r_3$ are in $R^+$, and
$S(\theta)=\pmatrix{\cos \theta & \sin \theta\cr \sin \theta &-\cos
\theta\cr}$, and $\alpha$, $\alpha'$, $\beta$, $\beta'$, $\gamma$, $\gamma'$
satisfy
 $$\alpha+\alpha'=\beta+\beta'=\gamma+\gamma'$$.

\vskip .2cm
\noindent
{\it Proof.} If $A$ and $A+J_o$ are both in ${\cal S}_8$, then the matrix
$AJ_o+J_oA$ is proportional to identity. This gives a set of linear equations
whose solutions can be obtained without difficulty
to yield
$$A=\pmatrix{
      0& a_{12}& a_{13}& a_{14}& a_{15}& a_{16}& a_{17}& a_{18}\cr
-a_{12}&      0& a_{14}&-a_{13}& a_{16}&-a_{15}& a_{18}&-a_{17}\cr
-a_{13}&-a_{14}&      0& a_{12}& a_{35}& a_{36}& a_{37}& a_{38}\cr
-a_{14}& a_{13}&-a_{12}&      0& a_{36}&-a_{35}& a_{38}&-a_{37}\cr
-a_{15}&-a_{16}&-a_{35}&-a_{36}&      0& a_{12}& a_{57}& a_{58}\cr
-a_{16}& a_{15}&-a_{36}& a_{35}&-a_{12}&      0& a_{58}&-a_{57}\cr
-a_{17}&-a_{18}&-a_{37}&-a_{38}&-a_{57}&-a_{58}&      0& a_{12}\cr
-a_{18}& a_{17}&-a_{38}& a_{37}&-a_{58}& a_{57}&-a_{12}&      0\cr}$$
 Then the requirement that the diagonal entries in $A^2$ be
equal to each other give the following equations after some
algebraic manipulations.
$$a_{13}^2+a_{14}^2=a_{57}^2+a_{58}^2$$
$$a_{15}^2+a_{16}^2=a_{37}^2+a_{38}^2$$
$$a_{17}^2+a_{18}^2=a_{35}^2+a_{36}^2$$
Thus we can parametrize $A$ by
$$a_{13}=r_1\cos \alpha,\quad a_{15}=r_2\cos\beta\quad a_{17}=r_3\cos \gamma$$
$$a_{14}=r_1\sin \alpha,\quad a_{16}=r_2\sin\beta\quad a_{18}=r_3\sin \gamma$$
$$a_{57}=r_1\cos \alpha',\quad a_{37}=r_2\cos\beta'\quad
  a_{35}=r_3\cos \gamma'$$
$$a_{58}=r_1\sin \alpha',\quad a_{38}=r_2\sin\beta'\quad
  a_{36}=r_3\sin \gamma'$$

Finally the requirement that the off diagonal terms in $A^2$ be equal to zero
gives quadratic equations, which can be rearranged and using trigonometric
identities they give
$\alpha+\alpha'=\beta+\beta'=\gamma+\gamma'$. \quad e.o.p.

Thus the set of matrices $A\in {\cal S}_8$ such that
$(A+J_o)\in {\cal S}_8$ constitutes
an eight parameter family and  the
equations of CDFN  correspond to the case $\alpha'+\alpha = \beta'+\beta =
\gamma'+\gamma=0 $.
Thus we have an invariant description of these equations, that we repeat here
for convenience.
\begin{eqnarray}
&F_{12}-F_{34}=0\quad F_{12}-F_{56}=0\quad F_{12}-F_{78}=0\cr
&F_{13}+F_{24}=0\quad F_{13}-F_{57}=0\quad F_{13}+F_{68}=0\cr
&F_{14}-F_{23}=0\quad F_{14}+F_{67}=0\quad F_{14}+F_{58}=0\cr
&F_{15}+F_{26}=0\quad F_{15}+F_{37}=0\quad F_{15}-F_{48}=0\cr
&F_{16}-F_{25}=0\quad F_{16}-F_{38}=0\quad F_{16}-F_{47}=0\cr
&F_{17}+F_{28}=0\quad F_{17}-F_{35}=0\quad F_{17}+F_{46}=0\cr
&F_{18}-F_{27}=0\quad F_{18}+F_{36}=0\quad F_{18}+F_{45}=0\cr \nonumber
\end{eqnarray}
The (skew-symmetric) matrix of such a 2-form is
$$ \left ( \begin{array}{cccccccc}
0& F_{12}&F_{13}&F_{14}&F_{15}&F_{16}&F_{17}&F_{18}\\
 &  0&F_{14}& -F_{13}&F_{16 }&-F_{15}&F_{18}&-F_{17}\\
 & &  0 &F_{12}& F_{17}&-F_{18}&-F_{15}&F_{16}\\
 & & & 0& -F_{18}&-F_{17}&F_{16}&F_{15}\\
 & & & & 0&F_{12}&F_{13}&-F_{14}\\
 & & & & & 0&-F_{14}&-F_{13}\\
 & & & & & & 0 & F_{12}\\
 & & & & & & &  0 \end{array} \right ) $$
We will  refer to the plane consisting of these forms as the
CDFN-plane. Let us now consider as the reference form
$J=\pmatrix{0&I \cr -I&0 \cr }$ instead of $J_o$.
$J$ can be obtained from $J_o$ by conjugation $ J = P^t J_o P$ with
$$P= \left ( \begin{array}{cccccccc}
1&0&0&0&0&0&0&0\\
0&0&0&0&1&0&0&0\\
0&1&0&0&0&0&0&0\\
0&0&0&0&0&1&0&0\\
0&0&1&0&0&0&0&0\\
0&0&0&0&0&0&1&0\\
0&0&0&1&0&0&0&0\\
0&0&0&0&0&0&0&1 \end{array} \right ) $$
Then the conjugation of the CDFN-plane by $P$ is given by the following
remarkable ( D=8 self-dual) 2-form
$$ F_{12} J + \pmatrix{ \Omega'& \Omega''\cr - {\Omega''}^t & -\Omega'\cr}$$
where $\Omega'$ is a D=4 self-dual 2-form and $\Omega''$ is a D=4
anti-self-dual 2-form.
\newpage
\vskip .5cm
\noindent
{\bf 4. The Geometry of ${\cal S}_{4k+2}$ }
\vskip .2cm
We prove that for odd n there are no linear subspaces
other than the one dimensional one.
\proclaim Proposition 6. Let ${\cal M}=\{A\in {\cal S}\mid (A+J_o)\in {\cal
S}\}$. Then ${\cal M}= \{ k J | k \epsilon {\bf R} \}$ for odd $n$.

\vskip .2cm
\noindent
{\it Proof.}
Let $A=\pmatrix{A_{11}&A_{12}\cr
-A_{12}^t&A_{22}\cr}$, where $A_{11}+A_{11}^t=0$, $A_{22}+A_{22}^t=0$.
As before if $(A+J_o)\in {\cal S}$ then $AJ_o+J_oA$ is
proportional to the identity.
This gives
$A_{11}+A_{22}=0$ and the symmetric part of $A_{12}$ is proportional to
identity. Therefore $A=kJ_o+\pmatrix{A_{11}&A_{12o}\cr A_{12o}&-A_{11}\cr}$,
where $A_{12o}$ denotes the antisymmetric part of $A_{12}$ and $k$ is a
constant. Then the requirement that $A\in {\cal S}$ gives
$$[A_{11},A_{12o}]=0,\quad A_{11}^2+A_{12o}^2+kI=0,\quad k\in R.$$
As $A_{11}$ and $A_{12o}$ commute, they can be simultaneously diagonalizable,
hence for odd $n$ they can be brought to the form
$$A_{11}=diag(\lambda_1\epsilon,\dots,\lambda_{(n-1)/2}\epsilon,0)$$
$$A_{12o}=diag(\mu_1\epsilon,\dots,\mu_{(n-1)/2}\epsilon,0)$$
where $\epsilon=\pmatrix{0&1\cr-1&0\cr}$, and $0$ denotes a $1\times
1$ block, up the the permutation of the blocks. If the blocks occur as shown,
clearly $A_{11}^2+A_{12o}^2$ cannot be proportional to identity.
It can also be
seen that except for $\lambda_i = \mu_i =0$
the same result holds for any permutation of the blocks.
\vskip 5mm
\noindent {\bf 5. Conclusion}
\vskip 2mm
To conclude we would like to emphasise that the choice of a linear
subspace of ${\cal S}_{2n}$ is incidental. Instead, one should try to
understand
the totality of the non-linear space of self-dual 2-forms. In that respect
the approach to self-duality given above might be a good starting point.

\newpage
\noindent {\bf References}
\vskip 5mm
\begin{description}
\item{[1]} A.H.Bilge, T.Dereli, \c{S}.Ko\c{c}ak,
``Self-dual Yang-Mills fields in
eight dimensions'' {\it Lett.Math.Phys.} (to appear)
\item {[2]}   E.Corrigan, C.Devchand, D.B.Fairlie and J.Nuyts, ``First-order
equations for gauge fields in spaces of dimension greater than four'', {\it
Nuclear Physics} {\bf B214}, 452-464, (1983).
\item{[3]}  B.Grossman, T.W.Kephart, J.D.Stasheff, ``Solutions to Yang-Mills
field equations in eight dimensions and the last Hopf map'', {\it Commun. Math.
Phys.}, {\bf 96}, 431-437, (1984).
\item{[4]} N.Steenrod, {\bf The Topology of Fibre Bundles} (Princeton U.P. ,
1951)
\item{[5]} S.Kobayashi,K.Nomizu, {\bf Foundations of Differential Geometry}
Vol.II ( Interscience , 1969)
\item{[6]} F.W.Warner, {\bf Foundations of Differentiable Manifolds and Lie
Groups}
(Scott and Foresman, 1971)
\end{description}
\end{document}